\documentclass[aps,pre,floats,showpacs,preprint]{revtex4}

\usepackage{graphicx}

\begin{document}

\title{Two-surface wave decay: improved analytical theory and effects on electron acceleration}
\author{A. Macchi\email{macchi@df.unipi.it}}
\author{M. Battaglini}
\author{F. Cattani}
\author{F. Cornolti} 
\affiliation{Istituto Nazionale per la Fisica della Materia (INFM), sezione A, 
Dipartimento di Fisica ``Enrico Fermi'', Universit\`a di Pisa, 
Via Buonarroti 2, 56127 Pisa, Italy}

\begin{abstract}
Two-surface wave decay (TSWD), i.e. the parametric excitation of electron surface waves, was recently proposed as an absorption mechanism in the interaction of ultrashort, intense laser pulses with solid targets. We present an extension of the fluid theory of TSWD to a warm plasma which treats boundary effects consistently. We also present test-particle simulations showing localized enhancement of electron acceleration by TSWD fields; this effect leads to a modulation of the current density entering into the target and may seed current filamentation instabilities.
\end{abstract}


\maketitle

\section{Introduction}
\label{intro}
The excitation of electron surface waves (ESWs) is a possible route to 
collisionless energy absorption in plasmas produced by the interaction
of sub-picosecond, high intensity (typically $\geq 10^{16}~\mbox{W/cm}^2$)
laser pulses with solid targets.
This regime is relevant for applications such as 
generation of ultrashort X-ray pulses, 
either as uncoherent thermal emission (\cite{gibbon} and references therein)
or high laser harmonics \cite{vonderlinde}, or production of energetic
electrons and ions \cite{gibbon,amiranoff,pukhov,umstadter}. 

Linear mode conversion of the laser wave into an ESW 
is not possible at a planar plasma-vacuum interface,
because the phase matching between the ESW and the incident wave
is not allowed;
the process can take place for 
specially tailored density profiles, e.g. grating 
targets with a density profile modulated
at the surface with a wavevector $k_p$. 
In this case, the condition for
the excitation of a ESW with wavevector $k_s$ 
(parallel to the plasma surface)
and the same frequency $\omega_L$ of the laser pulse is
\begin{equation}
k_L\sin\theta=k_s+k_p , \label{eq:pm}
\end{equation}
where $k_L=\omega_L/c$, $\theta$ is the angle of incidence of the laser
pulse.
Experimental investigations of laser absorption and X-ray emission
in grating targets are described, e.g., in Refs.\cite{gauthier,bastiani}.

{\em Nonlinear} mode conversion, i.e. parametric excitation 
of {\em two} ESWs is possible also in a simple ``step'' profile.
We name this process ``two--surface wave decay'' (TSWD) \cite{Macchi02}.
Examples of similar TSWD processes were previously considered 
in regimes other than intense laser--plasma interactions and in 
the electrostatic limit only \cite{gradov,stenflo}.
The general phase matching conditions for such a three--wave
process are given by
\begin{equation}
k_0=k_{+}+k_{-}~,~~\omega_0=\omega_{+}+\omega_{-},
\label{eq:MC}
\end{equation}
where $k_0,\omega_0$ and $k_{\pm},\omega_{\pm}$
are the wavevector and frequency of the
``pump'' wave and the two ESWs, respectively. 
If the electric field of the laser pulse 
acts as a ``pump'' for TSWD, then
$\omega_0=\omega_L$, $k_0=k_L\sin\theta$, and
equation (\ref{eq:MC}) implies that
two {\em sub--harmonic} ESW are generated with frequency around
$\omega/2$ (``$\omega \rightarrow \omega/2+\omega/2$'' TSWD).
However, also the ${\bf v}\times{\bf B}$ term of the Lorentz force
may excite TSWD. This was observed in particle-in-cell (PIC) simulations 
\cite{Macchi01} for normal laser incidence;
in this case, the ${\bf v}\times{\bf B}$ force drives 1D
oscillations at the plasma surface with frequency $2\omega_L$; after a 
few laser cycles the overlap of a {\em standing} wave with frequency $\omega_L$
was observed. This is a clear signature of a 
``$2\omega\rightarrow\omega+\omega$'' process
leading to two counterpropagating ESWs both having the
frequency of the laser pulse. 

According to the theoretical model \cite{Macchi02,MacchiLPB02},
the maximum growth rate of the $2\omega\rightarrow\omega+\omega$ case 
is found for normal incidence.
As said above, TSWD does {\em not} need a structured
target; however, it is worth noticing that in a grating target the
wavevector $k_p$ of the surface modulation [eq.(\ref{eq:pm})] equals the
wavevector of the two ESWs excited by the 
$2\omega\rightarrow\omega+\omega$ process at $\theta=0$.
Hence, TSWD is enhanced in such a grating target. 
It is then interesting to notice that
PIC simulations reported in Ref.\cite{bastiani}, 
showing the generation of two countepropagating ESWs in grating targets 
at $10^{16}~\mbox{W/cm}^2$, can now be interpreted as an evidence
of TSWD seeded by the surface grating.

At very high intensities ($\geq 10^{18}~\mbox{W/cm}^2$), 
simulations \cite{Macchi01}
show that TSWD enters a strongly nonlinear regime leading to 
strong rippling of the plasma surface. This might be relevant
to the surface instabilities
which have been observed in experiments \cite{Tarasevitch}
and play a detrimental role in high harmonic generation 
from solid targets \cite{norreys,zepf,tarasevitch2};  
such instabilities appear to grow even for
pulses of few tens of fs \cite{tarasevitch2} and thus must be of electronic
nature, since ions do not move on such a fast scale. Simulations also show
that nonlinear TSWD affects fast electron generation. This effect
is investigated in section \ref{sec:acceleration} by means of
test-particle simulations. Before that, in section \ref{sec:theory}
we present an improvement of the model of Ref.\cite{Macchi02} where
temperature effects have been included. In both cases, we restrict
ourselves to the case of the $2\omega\rightarrow\omega+\omega$ TSWD
at normal laser incidence, and we neglect collisions
for simplicity and because collective processes
dominate absorption at intensities $\geq 10^{16}~\mbox{W/cm}^2$.

\section{Temperature and boundary effects on TSWD}
\label{sec:theory}
In Ref.\cite{Macchi02} the TSWD growth rate was calculated for a 
step-like density profile [$n_i=n_0\Theta(x)$, being $\Theta(x)$ 
the Heaviside step function] using
an Eulerian, fluid model with immobile ions
and using the cold plasma approximation. 
We adopted the following expansion for all fields
\begin{eqnarray}
{f}(x,y,t)&=&{f}_i(x,t-y\sin\theta/c)+
              \epsilon f_{0}(x)e^{i\omega_0(t-y\sin\theta/c)}+ \nonumber \\ 
          & & \epsilon^2[f_{+}(x)e^{ik_{+}y-i\omega_{+}t}+
                         f_{-}(x)e^{ik_{-}y-i\omega_{-}t}]
,
\label{eq:expansion}
\end{eqnarray}
where $\epsilon$ is a small expansion parameter, and
$f$ stands for either the electron density or velocity
or for the EM fields in the $(x,y)$ plane.
The first term ($f_i$) of eq.(\ref{eq:expansion}) includes zero-order, 
unperturbed fields or 
oscillating fields that are non-resonant with the excited modes;
the term $f_0$ represents the ``pump'' field 
at the frequency $\omega_0$;
the last term is the sum of two counterpropagating surface modes. 
For the $2\omega\rightarrow\omega+\omega$ process at $\theta=0$,
$\omega_0=2\omega_L$ and $k_{+}=-k_{-}$, and the zero-- and first--order
fields do not depend on $y$ (i.e. they are ``1D'' fields).
The coupling between the pump and the surface modes (of order $\epsilon^3$)
originates from the nonlinear terms in the Euler equation
($-en_e{\bf v}\times{\bf B}$, $-m_en_e{\bf v}\nabla{\bf v}$) and the
current density ${\bf J}=-en_e{\bf v}$.

The calculation for a warm, isothermal plasma 
proceeds very similarly \cite{battaglini} 
and thus only the differences from the ``cold'' case and their 
consequences are discussed below, while details of the calculation
will be published elsewhere. The only difference in the starting
Maxwell-Euler systems of equations comes from the pressure term
in the Euler equation for electrons:
\begin{equation}
m_e d{\bf v}/dt=-{e}n_e({\bf E}+{\bf v}\times{\bf B})-k_BT_e{\nabla}n_e ,
\label{eq:euler}
\end{equation}
where $n_e$, ${\bf v}$ and $T_e$ are the density, fluid velocity and 
temperature of electrons.

Assuming reflective conditions at the plasma boundary ($x=0$),
the charge density $\rho=e(n_i-n_e) \neq 0$ only in a surface
layer with a thickness on the order of the Debye length. At $T_e=0$ this 
corresponds to a surface charge layer [i.e. $\rho \sim \delta(x)$, where
$\delta(x)$ is the Dirac delta function], while the longitudinal 
velocity ($v_x$) is discontinous at the surface. This is relevant because
in the $T_e=0$ case nonlinear coupling terms involving the product of 
fields that are singular at $x=0$ occur. The pressure term removes such 
singularities
and enables to evaluate all such terms correctly. Therefore, by
calculating the TSWD growth rate and {\em then} taking the 
$T_e \rightarrow 0$ limit 
one achieves an improvement of the result obtained in the cold plasma case.

The final result of the calculation is the $2\omega\rightarrow\omega+\omega$
growth rate $\Gamma$ shown in Fig.\ref{fig:rate} (the complete 
analytical expression is very lengthy \cite{battaglini} 
and is not reported here).
With respect to the $T_e=0$ case, the two divergences of $\gamma$
are both quenched for growing $T_e$, but for different
reasons. The divergence
for $\omega\rightarrow\omega_p/\sqrt{2}$ is quenched
since in this limit the ESWs have shorter wavelengths and are thus
more affected by the thermal pressure that inhibits the formation
of small-scale structures. 
The ``pump'' resonance at $\omega\simeq\omega_p/2$, due to the 
excitation of plasmons by the ${\bf v}\times{\bf B}$
force, is quenched by energy transport out of the skin layer 
because for $T_e \neq 0$ the 
resonant plasmon propagates into the overdense plasma.

\section{Electron acceleration}
\label{sec:acceleration}
At high intensity of the laser pulse, 
it is well known that most of the absorbed energy
goes into ``fast'' electrons injected into the overdense
plasma region. At normal incidence, due to the leading
action of the ${\bf v}\times{\bf B}$ force, 
fast electron bunches are produced twice per laser cycle.
The PIC simulations of Ref.\cite{Macchi01}
showed that after the growth of nonlinear TSWD the generation of 
fast electrons was enhanced near spatial maxima of the standing surface 
oscillation (Fig.\ref{fig:phasespace}, top). Thus,
this effect may give an ``imprint'' for the formation of electron 
filaments, whose size and spacing would be close to (and scaling with)
the laser wavelength as observed in other simulations 
\cite{lasinski,sentoku},
and affect energy transport by fast electrons into the plasma.

To investigate this effect further, we performed ``test particle'' 
simulations of electron motion in the overlapping ``pump'' and 
TSWD fields. In other words, we solve the equation of motion for 
electrons into a force field of the form given by eq.(\ref{eq:expansion}),
i.e. the sum of an one-dimensional (1D) force of frequency $2\omega$ 
(whose analytical expression is obtained from the theoretical model) 
and the force field from a standing 2D surface wave of frequency $\omega$.
A similar study, focused on the acceleration of electrons by a
single ESW, was recently reported \cite{riconda}.

The 1D and 2D fields vary in time
as $\cos 2\omega t$ and $\sin\omega t$, so that the temporal maxima and minima
of the ESW field are always coincident in time with maxima of the 1D field, 
as it was found in theory and simulation \cite{Macchi02,Macchi01}.
In what follows, $x$ is the direction normal to the plasma surface
located at $x=0$ (the plasma occupies the $x>0$ region) and $y$ is the
direction of propagation of ESW.  
The field amplitudes 
are chosen so that the system remains far from relativistic conditions 
and the ESW field can be considered as a perturbation. The ratio between the 
amplitudes of standing SW and the $2\omega$ pump was varied as a parameter.
For the simulations reported in this paper, in terms of 
normalized amplitudes, $a=eE/m_ec\omega$ where $E$ is the amplitude of the 
electric field, the values are 
$a_L = 0.2$ for the laser field and $a_{ESW} = 0.019$ for the ESW field. 
The wavevectors of the laser wave and of the ESW are given by the 
known expressions for a cold plasma with density 
$n_e/n_c=\omega_p^2/\omega^2=5$.
To initialize the simulations, we gave the particles an initial position
$x(0)>0$ such that the evanescent fields are negligible at that point,
and an initial velocity $v_x(0)<0$ (on average $v_x=-0.1 c$) 
so that the particles move towards
the surface region, where the ESW fields are localized. The particles 
are distributed in $y$ uniformly over a region of width $\lambda_s$,
where $\lambda_s$ is the ESW wavelength. 

The ($y,p_x$) projection of the test particle phase space 
(Fig.\ref{fig:phasespace}, bottom) 
looks similar to the one from the fully self-consistent PIC simulation 
(Fig.\ref{fig:phasespace}, top), 
showing that electron acceleration is indeed enhanced 
near spatial maxima of the 
standing wave. The overlap of the SW fields with the 
${\bf v}\times{\bf B}$-driven fields (homogeneous along the $y$-direction)
thus leads to a modulation along $y$ of the longitudinal momenta. 

The enhancement (or quenching) of the accelerating field occurs once per
laser cycle (because the ESW frequency equals the laser frequency), 
but with opposite phase 
between contiguous maxima of the standing ESW.
This is shown by Fig.\ref{fig:fingers} 
where the complete ($x,p_x$) phase space projection, showing
electron bunches generated at $2\omega$ rate, is compared 
with the same plot but including now only the electrons 
whose starting position lies within an interval of $\lambda/4$ width 
around the $y=\lambda/4$ spatial maximum of the ESW.
It is thus evident how the most energetic electrons in the bunches 
penetrating in the $x>0$ region are generated near the ESW maximum and
at $\omega$ rate, i.e. with the ESW frequency. Taking only the electrons
around the $y=3\lambda/4$ spatial maximum gives an almost identical picture,
except that the most energetic bunches are now out of phase 
by an angle of $2\pi/\omega$ with respect to those coming from
$y=\lambda/4$.

Although the amplitude of ESW fields is 0.1 times the 
pump field amplitude, the relative modulation of the longitudinal momentum 
$p_x$ is about 30\%. This can be qualitatively explained at follows.
The electrons acquire energy from an evanescent, oscillating field 
if their transit time across the region where the field has 
non-vanishing amplitude is shorter than an oscillation period; this is
the condition for non-adiabaticity of electron motion with respect
to the field. The ratio between the transit time and the oscillation 
period is approximately given by the parameter $\eta=L/v_0T$ where $L$ 
is the evanescence length, $v_0$ is the average velocity of the electron
and $T$ is the oscillation period. For the ESW, $T=2\pi/\omega$ and
$l_{ESW}=(c/\omega)\sqrt{\alpha-2}/(\alpha-1)$ 
(where $\alpha=\omega_p^2/\omega^2$)
while for the 1D field at $T=\pi/\omega$ and
$l_{2\omega}=(c/2\omega)(1/\sqrt{\alpha-1})$ 
and \cite{Macchi02}. Thus,
$\eta_{ESW}/\eta_{2\omega}=\sqrt{(\alpha-2)/(\alpha-1)}<1$ which means
that the electron motion (for a given $v_0$) is more non-adiabatic
with respect to the ESW field rather than to the $2\omega$ oscillation.
Thus, the contribution of the ESW in accelerating (or decelerating)
electrons is enhanced by the relatively low frequency and short
evanescence length. 

The ``imprint'' effect of the standing ESW is also noticeable in the
contourplot of the fluid velocity of electrons, shown in 
Fig.\ref{fig:flow}. The velocity field has been computed by averaging
the velocity over a spatial grid, as in PIC codes, and over time.
As a consequence the electric current in the overdense plasma region
has a spatial transverse modulation with the same wavelength of ESWs.
The simulations of 
Ref.\cite{bastiani} for ``grating'' targets also show a modulation of
the fast electron current correlated with same period of the grating, that
may be due to the local enhancement of the longitudinal field by TSWD
or to the geometrical ``funnel'' effect of the surface deformation
\cite{ruhl}. Our test particle results indeed show that due to TSWD a 
modulated fast electron current can be produced even by a plane--wave
pulse on a flat surface.

From figures \ref{fig:phasespace} and \ref{fig:fingers} 
it is also found that only near ESW spatial maxima 
a few electrons are ejected into vacuum, i.e. in the $x<0$ region
(in Fig.\ref{fig:fingers}, for these electrons the oscillation of 
$p_x$ vs. $x$ is due to 
the effect of the ${\bf v}\times{\bf B}$ force in vacuum). Their origin is
likely to be due to the longitudinal field component directed into vacuum 
that is
associated to the ESWs. In Ref.\cite{riconda}, the features of electrons 
accelerated in vacuum by the field of a single ESW are investigated.
Note, however, that electrostatic back-holding fields
are not self-consistently included in test particle simulations and thus the
number of electrons escaping in vacuum is likely to be overestimated.
Nevertheless, it is interesting to notice that in the PIC simulations at very
high intensities \cite{Macchi01,MacchiLPB02}
``plumes'' of electrons extending into the vacuum region 
are generated near the maxima of the standing ESW.
A ``plume'' structure in the vacuum region is also evident in the 
velocity field shown in Fig.\ref{fig:flow}.

\subsubsection*{Acknowledgments.} This work was partly supported 
by the Italian Ministery of University and Research (MIUR) 
through the project
``Generation of fast electron and ion beams by superintense laser
irradiation''.
Discussions with Serena Bastiani-Ceccotti, Caterina Riconda 
and Francesco Pegoraro are gratefully acknowledged.

\newpage

\begin{figure}
\caption{Growth rate of the $2\omega\rightarrow\omega+\omega$ process 
vs. normalized electron density $n_e/n_c=\omega_p^2/\omega_L^2$ ($x$-axis) 
and temperature $\sqrt{k_BT_e/m_e c^2}$ (labels). 
The dashed line is the ``cold'' result previously obtained \cite{Macchi02}. 
The growth rate is normalized to $a_L^2\omega$ 
where $a_L=eE_L/m_ec\omega_L$ is the dimensionless
field amplitude of the laser pulse.}
\label{fig:rate}
\end{figure}

\begin{figure}
\caption{Top: ($y,p_x$) phase space projections from PIC simulations 
\cite{Macchi01} at two subsequent times, showing electron acceleration localized near spatial maxima of the growing surface oscillations. Bottom: ($y,p_x$) phase space projection from test particle simulations, showing similar features. The ``black stripe'' around $v_x=-0.1$ represents electrons that have not been reached yet the surface region at the time shown.}
\label{fig:phasespace}
\end{figure}

\begin{figure}
\caption{Left: phase space ($p_x$,$x$) projection (integrated over $y$) 
from test particle simulations, showing fast electron jets penetrating 
into the plasma region ($x>0$) where they move ballistically ($p_x \propto x$).
The jets are produced twice per laser cycle. 
All particles in the simulation are included in the plot.
Right: same as left side, but restricted 
to particles around $y=\lambda_s/4$ (spatial maximum of the standing 
surface wave), showing enhanced acceleration around this position once 
per laser cycle. In both figures electrons propagating into vacuum ($x<0$) 
are also evident. They are found only near ESW maxima ($y \simeq \lambda_s/4$, 
$y \simeq 3\lambda_s/4$).}
\label{fig:fingers}
\end{figure}

\begin{figure}
\caption{Contours of time-averaged fluid velocity from test particle
simulations, showing a transverse modulation due to the enhanced 
electron acceleration near ESW maxima ($y\simeq \lambda_s/4$, 
$y \simeq 3\lambda_s/4$).}
\label{fig:flow}
\end{figure}

\end{document}